\documentclass[pra,aps,twocolumn,eqsecnum,showpacs]{revtex4}
\usepackage{mathtext}
\usepackage[T2A]{fontenc}
\usepackage[cp1251]{inputenc}
\usepackage[english]{babel}
\usepackage{epsf}
\usepackage{psfrag}
\usepackage{graphicx}
\usepackage{amssymb,amsmath}

\begin{document}

\title{Reflection of resonant light from a plane surface of an ensemble of motionless point scatters: quantum microscopic approach}
\author{A.S. Kuraptsev${}^{1}$ and I.M. Sokolov${}^{1,2}$
\\
{\small $^{1}$Peter the Great St.Petersburg Polytechnic University, 195251, St.-Petersburg, Russia }\\
{\small $^{2}$Institute of Analytical Instrumentation, Russian Academy of
Sciences, 198103, St.-Petersburg, Russia }\\
}
\date{\today}

\sloppy



\begin{abstract}
On the basis of general theoretical results developed previously in [JETP 112, 246 (2011)], we analyze the reflection of quasi-resonant light from a plane surface of dense and disordered ensemble of motionless point scatters. Angle distribution of the scattered light is calculated both for s- and p- polarizations of the probe radiation. The ratio between coherent and incoherent (diffuse) components of scattered light is calculated. We analyze the contributions of scatters located at different distances from the surface and determine on this background the thickness of surface layer responsible for reflected beam generation. The inhomogeneity of dipole-dipole interaction near the surface is discussed. We study also dependence of total reflected light power on the incidence angle and compare the results of microscopic approach with predictions of Fresnel reflection theory.  The calculations are performed for different densities of scatters and different frequencies of a probe radiation.
\end{abstract}
\maketitle
\section{Introduction}

The vast majority of experimental optical detection methods are based on analysis of radiation scattered from investigated medium. Among these methods ones based on measurements of coherent component of scattered radiation have a range of advantages. Reflection of light from resonant media have even formed a special area of optics (see \cite{1}-\cite{9g} and references therein). Among a great variety of resonant media the disordered ensembles of point-like scatters which motion can be neglected  take a special place. The physical model of motionless scatterers is commonly used for description of interaction between impurity centers in solids and electromagnetic radiation. It also can be used for description of cold atomic ensembles prepared in special atomic trap.

Dense ensembles in which the average interatomic distance and the mean free path of photon are comparable with the wave length of resonant radiation have attracted a special interest recently. It is connected with both exciting physical properties of such systems as its widespread practical application in quantum metrology, frequency standardization and quantum information science \cite{10}-\cite{18}.

The interaction of resonant light with dense ensemble has a range of important features which are usually neglected in case of dilute ensembles. If the average interatomic distance is comparable with resonant wavelength the atoms can not be considered as independent scatters of electromagnetic waves (hereafter we will associate point scatters with atoms for brevity). In this case we deal with so-called cooperative scattering \cite{19},\cite{20}. Interatomic dipole-dipole interaction significantly influences on the optical characteristics of a medium. Collective effects cause density-dependent shifts of atomic transition as well as distortion of spectral line shape \cite{21},\cite{22}. The real part of dielectric permittivity of dense atomic ensemble can be negative in some spectral area \cite{23}-\cite{27}.

Modification of optical characteristics caused by dipole-dipole interaction manifests itself differently for spatial areas inside the medium and near its surface. The atoms located in the subsurface layer interact predominantly with atoms situated at
one side of them, inside the ensemble. The surface layer is generated. Its width depends on density of the sample and is about 1.5-2 inverse wave numbers \cite{25}. This subsurface region significantly influences both the incoherent scattering and coherent reflection.

The goal of this paper is to study the reflection of quasi-resonant light from the plane boundary between vacuum and dense ensemble of point scatters. We analyze the influences of the features of dipole-dipole interaction caused by the subsurface spatial inhomogeneity on the properties of reflection.

Inhomogeneity of the optical properties and spatial disorder of atoms in the ensemble restrict the classical description using Fresnel equations which require the mean free path of photon and the wavelength of probe radiation much greater than the average interatomic distance. In this paper we use the consequent quantum microscopic approach \cite{19}. This approach allows us to obtain both coherent and incoherent (diffuse) component of scattered light. Note that nearly all previous applications of the method developed in \cite{19} were devoted to analysis of incoherent scattering of the light from random media, particularly for study of multiple recurrent scattering. This analysis requires calculation of average intensity of scattered light. In the present paper for the first time we use this approach for description of coherent mirror scattering from random media. We calculate mean electric field of the light scattered by disordered media with sharp boundary as a sum of individual contributions of all atoms. Such approach allows us to analyze the partial contributions of the layers of the medium located at different distances from the surface and analyze the properties of reflected wave depending on density of the atoms, frequency of the probe light, its polarization and on angle of incidence.

\section{Basic assumptions and approach}
The calculation of resonant reflection in this paper will be made on the basis of a microscopic quantum approach developed in \cite{19}. This approach is based on the non-stationary Schrodinger equation for the wave function of the  joint system consisting of $N$ motionless atoms and electromagnetic field. All atoms are assumed to be identical and have a ground state $J = 0$ separated by the frequency $\omega_{0}$ from an excited $J = 1$ state. The excited state has the Zeeman structure so there are three sublevels for each atom which differ by the value of angular momentum projection $m=-1,0,1$. Such structure of atomic levels allows us to describe the effects connected with vector nature of electromagnetic field in case of dense medium correctly. Note that the standard two-level scalar model does not have this advantage. For detailed comparison of these models  see \cite{34}, \cite{35}.

The Hamiltonian of the joined system $H$ can be presented as a sum of two operators $H_{0}+V$. Here $H_{0}$ is the sum of the Hamiltonians of the free atoms and the free field, $V$  is the operator of their interaction. The wave function is found as an expansion in a set of eigenstates $\{|l\rangle\}$ of the operator $H_{0}$.

The key simplification of the approach  is in restriction of the total number of states $|l\rangle$ taken into account. We consider only the vacuum state $\psi_{g'}$ (all atoms are in the ground state and there is no photon), the one-fold excited atomic states $\psi_{e_{a}^{m}}$ (one atom is excited and there is no photon), the one-fold excited states of field subsystem $\psi_{g}$ (there is one photon  and there are no excited atoms), and the nonresonant states $\psi_{e_{a}^{m}e_{b}^{m'}}$  with two excited atoms and one photon in the field subsystem. The complex index $e_{a}^{m}$ used here contains information about both the number of excited atom $a$ and the Zeeman sublevel which is populated.

In the rotating wave approximation it is enough to take into account only the states $\psi_{e_{a}^{m}}$ and $\psi_{g}$. States without excitation both in atomic and field subsystem $\psi_{g'}$ allow us to describe coherent states of the weak probe radiation \cite{26}. Nonresonant states with two excited atoms and one photon $\psi_{e_{a}^{m}e_{b}^{m'}}$ are necessary for a correct description of the dipole-dipole interaction at short interatomic distances.

The amplitude of state $\psi_{g'}$ does not change during the evolution of the system, because transitions to this state from other states taken into account are impossible. The transition from $\psi_{g'}$ to any other state is also impossible. The total set of equations for the other quantum amplitudes $b_{e_{a}^{m}}$, $b_{g}$ and $b_{e_{a}^{m}e_{b}^{m'}}$ is infinite because of the infinite number of field modes.  From this infinite set of equation we can pick out the finite subset of $3N$ algebraic equations for the Fourier component of amplitudes of one-fold atomic excitation $b_{e_{a}^{m}}$. Its formal solution can we written as follows:
\begin{equation}
b_{e_{a}^{m}}(\omega)=\sum_{b,m'}R_{e_{a}^{m}e_{b}^{m'}}(\omega)b_{e_{b}^{m'}}^{0}(\omega) \label{1}
\end{equation}
The matrix $R_{e_{a}^{m}e_{b}^{m'}}$ is the resolvent of considered system projected on the one-fold atomic excited states. This matrix describes the multiple photon exchange among atoms and it depends both on the spatial location of atoms and on the frequency of probe light (see \cite{19}, \cite{27}). The vector $b_{e_{b}^{m'}}^{0}(\omega)$ describes the interaction of atoms with external radiation which is is assumed to be a plane monochromatic wave. This vector depends on the direction of probe light and its polarization.
\begin{equation}
b_{e_{b}^{m'}}^{0}(\omega)=-\textbf{d}_{e_{b}^{m'};g_{b}}\textbf{e}/\hbar E_{0}\exp(i\textbf{k}_{0}\textbf{r}_{b})
\label{2}
\end{equation}
In this equation $\textbf{d}_{e_{b}^{m'};g_{b}}$ is dipole matrix element for transition from the ground g to the excited $m'$ state of atom b, $E_{0}$ is an amplitude of probe radiation, $\textbf{k}_{0}$ and $\textbf{e}$ are its wave vector and unit polarization vector, $\textbf{r}_{b}$ is the radius-vector of the atom b.

Microscopic approach allows us to consider atomic ensemble with arbitrary shape and spatial distribution of atomic density. In this paper we will consider sample in the form of a rectangular parallelepiped with random but uniform on average spatial distribution of atoms. The edge lengths of the parallelepiped are $l_x$, $l_y$ and $l_z$. The quantization axis $z$ is directed perpendicular to the front surface of an ensemble, the axis $x$ along the projection of the wave vector of the probe light on this surface. The angle between the direction of probe light propagation and the axis $z$ is $\theta_{0}$. The polarization of light is assumed to be linear. We will analyze two types of linear polarization: parallel to the plane of incidence (p- polarization) and perpendicular to this plane (s- polarization). The polarization vectors corresponding to these polarizations $\textbf{e}_{p}$ and $\textbf{e}_{s}$ can be presented in the basis of unit cyclic vectors $\textbf{e}_{-1}$, $\textbf{e}_{0}$, and $\textbf{e}_{+1}$ using standard relations \cite{28}. The advantage of this basis is that the dipole moment projections are given by simple equations $\textbf{d}_{e_{b}^{-1};g_{b}}\textbf{e}_{-1}=\textbf{d}_{e_{b}^{0};g_{b}}\textbf{e}_{0}=\textbf{d}_{e_{b}^{+1};g_{b}}\textbf{e}_{+1}=\sqrt{3\hbar c^{3}\gamma_{0}/4\omega_{0}^{3}}$ ($\gamma_{0}$ is the natural linewidth).

Describing the polarization properties of scattered light we will use the frame of reference connected with the direction of scattered light (with $z'$ axis along the wave vector) It is connected with typical arrangement of polarization measurement when polarization analyzers are oriented in accordance with the direction of detected radiation. So long as our paper is devoted mainly on the investigation of coherent scattering we will further restrict ourselves to the case when both incident and scattered waves have the same type of polarization (s- or p-).

Numerical calculation of the amplitudes $b_{e_{a}^{m}}(\omega)$  on the basis of (\ref{1})-(\ref{2}) allows us to obtain amplitudes of other quantum states $b_{g}$ and $b_{e_{a}^{m}e_{b}^{m'}}$. Thus we can obtain the wave function and any physical observable, particularly electric field strength of scattered light and light intensity (see \cite{19} for detail). It gives us opportunity to compare the results obtained in the framework of quantum microscopic approach with Fresnel equations.

The angular distribution of scattered light power contains a speckle component because of light interference from a big number of randomly distributed point scatters. In experiments the radiation averaged over an area of photodetector and integrated over definite time interval is measured. Therefore in our calculations we perform multiple averaging of results over random spatial configurations of atoms by Monte-Carlo method. To analyze the coherent component of scattered light we average the electric field strength and then we calculate intensity of this component. To calculate total light intensity we average light intensity itself. Note also that this averaging allows taking into account partly the residual thermal motion in cold atomic ensembles.

\section{Results and discussion}
\subsection{Angular distribution of scattered light}
We start with analysis of angular distribution of light scattered by optically dense plate. Figure 1 shows the power of light scattered in a unit spherical angle as a function of both polar angle [Fig. 1(a)] and azimuthal angle [Fig. 1(b)]. The calculation is performed for atomic ensemble with $l_{x}=110$, $l_{y}=55$, $l_{z}=6.53$ (hereafter in this paper we use the  inverse wave number of resonant light $k_{0}^{-1}$ as a unit of length, $k_{0}^{-1}=\lambda_{0}/2\pi$). The probe radiation is assumed to be exactly resonant with free atom transition, its frequency detuning is equal to zero $\Delta=\omega-\omega_{0}=0$, the angle of incidence is $\theta_{0}=17.5°$. As the computational difficulty increases rapidly with the number of atoms the atomic density is chosen not very big $n=0.05$ which corresponds to the mean free path of photon $l_{ph}=1.63$. However, as it was shown in \cite{27} the collective effects caused by dipole-dipole interaction play a significant role for such density.

Under considered conditions the reflection of light takes place not only from the front surface of the ensemble but from the side surfaces as well. To eliminate the influence of reflection from side surfaces we take into account only secondary radiation from atoms located sufficiently far from the sides, approximately 60\%  of the total number of atoms. In experiment such elimination can be performed by means of a diaphragm.
\begin{figure}
\begin{center}
{$\scalebox{0.4}{\includegraphics*{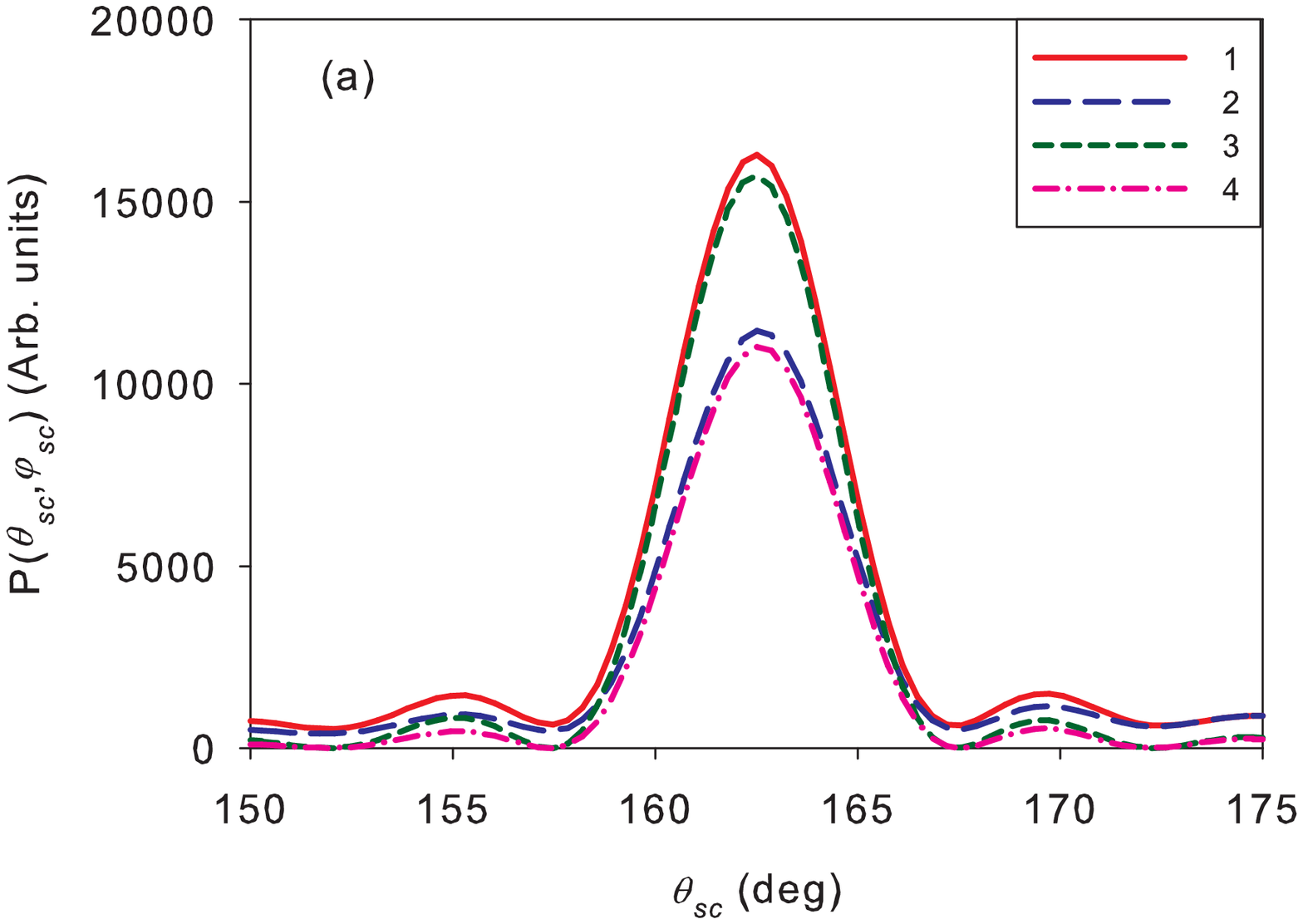}}$ }{$\scalebox{0.4}{%
\includegraphics*{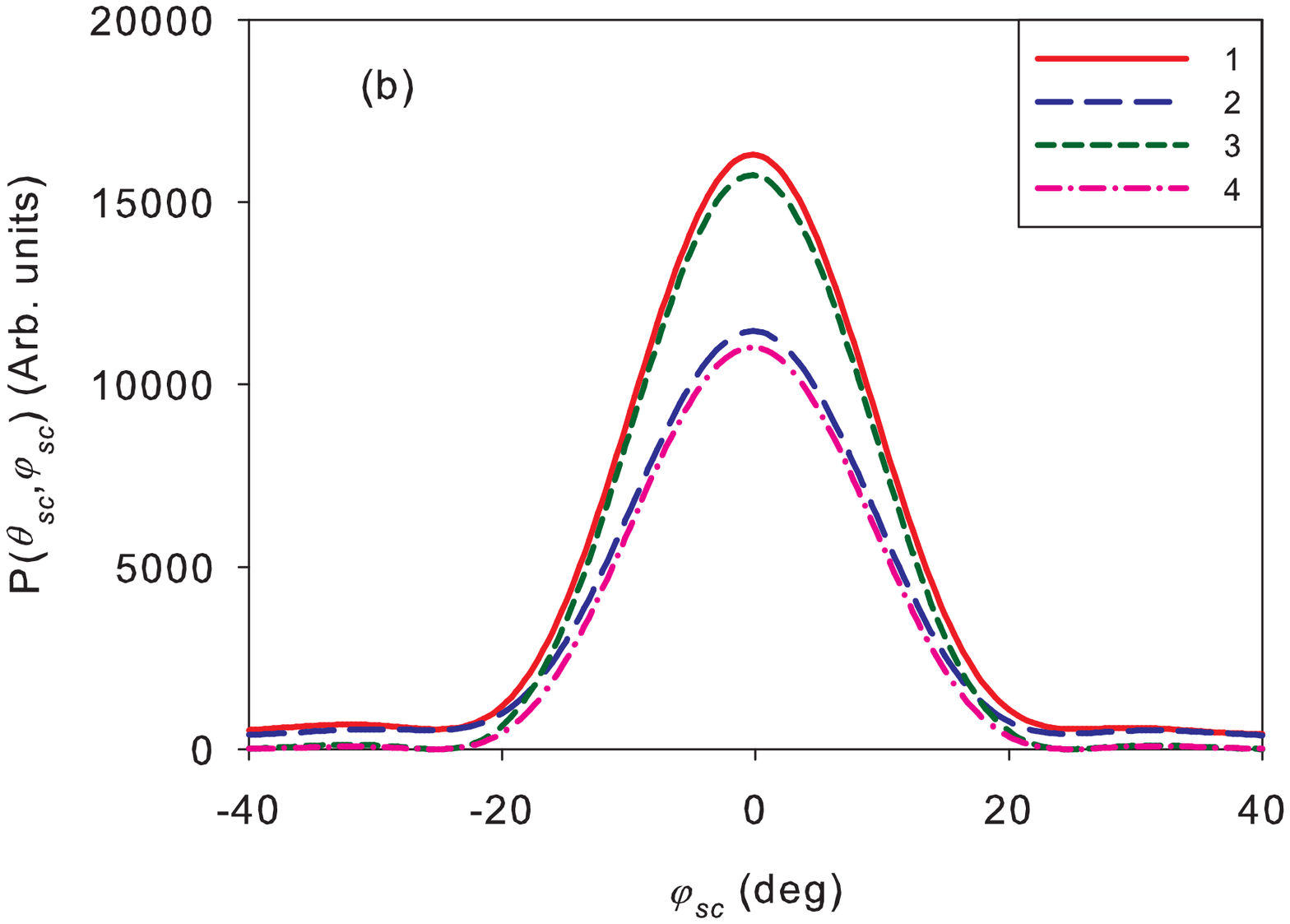}}$ }
\caption{(color online) Angle distribution of the scattered light power. (a) $\varphi_{sc}=0$, (b) $\theta_{sc}=\pi-\theta_{0}$. 1 s- polarization, 2 p- polarization, 3 s- coherent component, 4 p- coherent component}
\end{center}
\par
\label{fig1}
\end{figure}

Fig. 1(a) shows that the maximum of the scattered light power obtained in the frame of microscopic approach corresponds to a well known reflection law ($\theta_{sc}=\pi-\theta_{0}$, $\varphi_{sc}=0$). Besides of the main peak we observe weak satellite peaks caused by diffraction from the rectangular front surface and a small contribution caused by diffuse scattering. The height and width of the main peaks in the Fig. 1(a) and Fig. 1(b) are determined by the sizes $l_{x}$ and $l_{y}$ respectively. As we have $l_{x}>l_{y}$ the main peak of $\theta$-distribution (Fig. 1(a)) is more narrow than one of $\varphi$-distribution (Fig. 1(b)).

For comparison  we performed similar calculations for different sizes of atomic ensemble. The height of the main peak increased with size but its width decreased so that the total intensity of reflected light was proportional to the area $l_{x}l_{y}$.

In Fig.1 we included  the results of calculation of total scattered light power and its coherent component. The ratio of coherent component to total power exceeds 0.85 even for relatively small atomic density $n=0.05$. It confirms the fact that cooperative effects  play a significant role for this density.

Note, that there are several physically different cooperative effects which can take place under light interaction with dense and cold atomic ensembles.  Such phenomena as super-radiance, lasing in disordered media and Anderson localization attract great attention recently.
Physical effect studied in the present work has a bit different nature than all phenomena mentioned above. In our case collective effect does not assume multiple scattering. It is determined by interference of secondary radiation emitted by different atoms located in subsurface layer of the cloud. Similar interference causes coherent Rayleigh forward scattering which cross section is proportional to squared number of atoms in all ensemble.  In our case intensity of reflected light is proportional to squared amplitude of electric field and consequently to squared number of atoms in the mentioned subsurface layer. In the next subsection we will analyze formation of reflected beam in more details.

\subsection{Microscopic analysis of reflected beam generation}
Let us analyze now how the atoms located at different distances from the surface influense on the reflected signal generation. Our approach allows us to calculate the reflected signal taking into account only finite size layer near the front surface. Figure 2a shows corresponding results for three layers with the depth $d=0.5l_{ph}, l_{ph}, 2l_{ph}$ ($l_{ph}$ is the mean free path of photon obtained in \cite{24}) as well as for the whole ensemble with depth $7l_{ph}$. In case of the first and the second layer the power of reflected light exceeds the total reflected signal. For the third layer the result is less than one for a whole ensemble. Such behavior can be explained by the interference of the electromagnetic waves scattered by different atoms. When we consider relatively thin layers the phase increment on its thickness is small so the interference is constructive and the power of reflected signal increases with the depth of layer. If thickness of the layer becomes comparable with wavelength the dephasing of electromagnetic waves scattered by different atoms becomes important and the destructive influence of the interference has to be taken into account. It looks like an interference of light in transparent thin films. But in our case we deal with resonant atomic ensembles and the absorption is very important. Under considered conditions the mean free path of photon is approximately equal to quarter of the resonant light wavelength. Atoms located sufficiently far from the front surface do not influence on the coherent scattering, i.e. on reflection. It can be seen in the Fig. 2b. We observe saturation in dependence of reflected light power on the depth of layer. The curve in Fig.2b flattens out at depth of the layers greater than $(3.5\div 4)l_{ph}$. For bigger density the mean free path of photon is smaller and the saturation is observed for smaller depth.
\begin{figure}
\begin{center}
{$\scalebox{0.4}{\includegraphics*{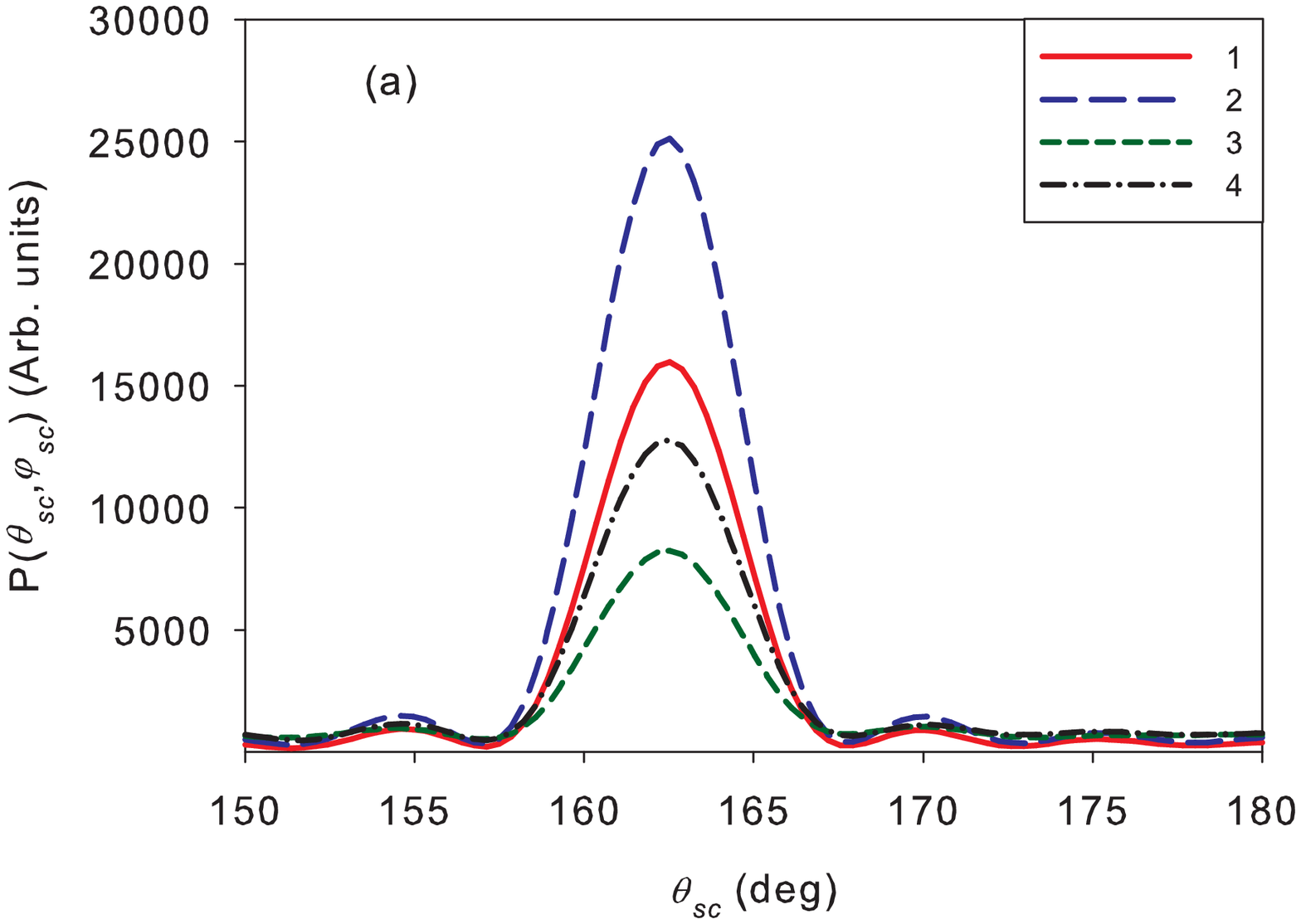}}$ }
{$\scalebox{0.44}{\includegraphics*{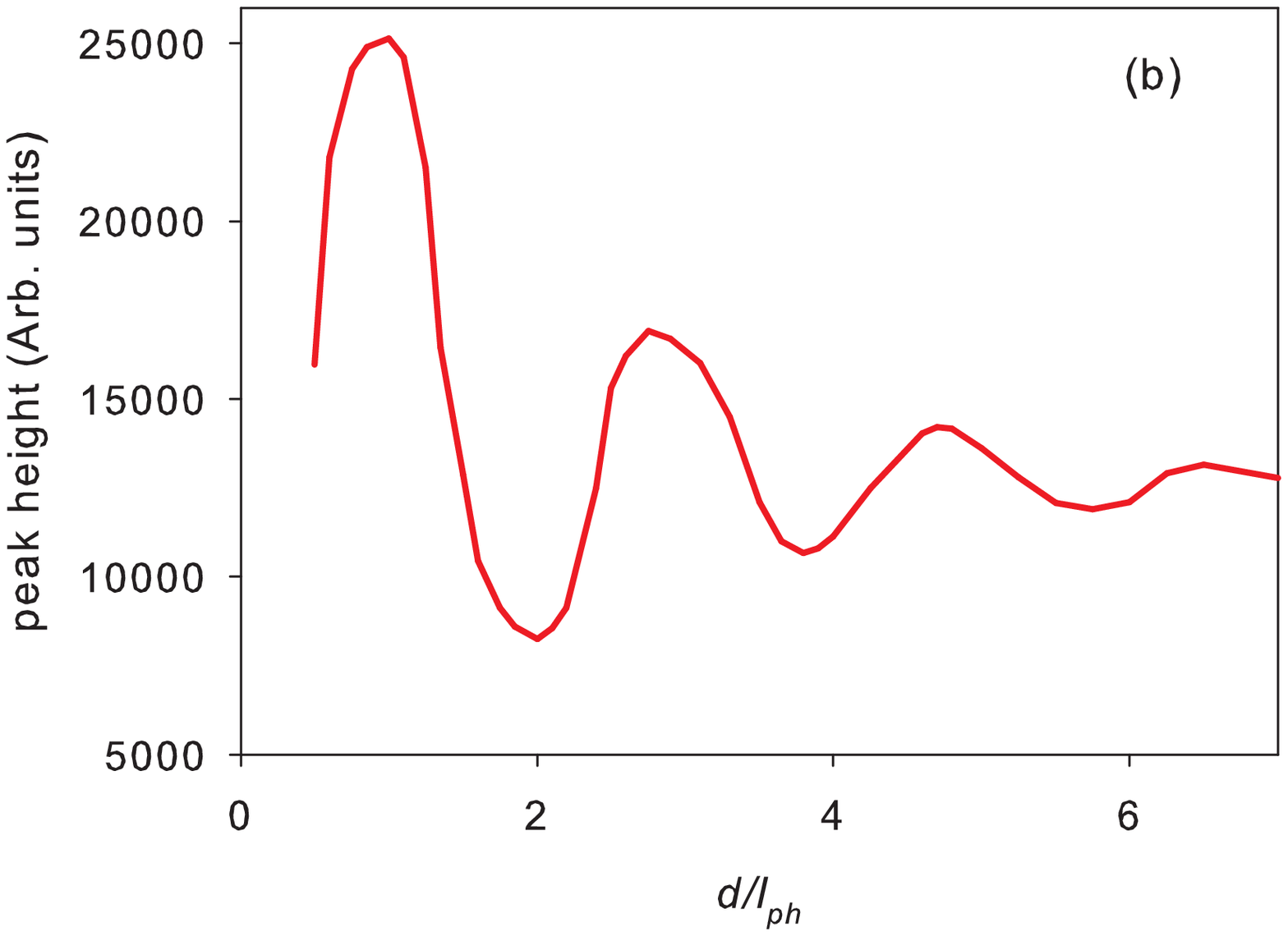}}$ }
\caption{(color online) a) Angle distribution of light scattered by subsurface layers; s- polarization; $\varphi_{sc}=0$;  depth of the layer: 1 - $0.5l_{ph}$, 2 - $l_{ph}$, 3 - $2l_{ph}$, 4 - $7l_{ph},$
b) Maximum of the function $P(\theta_{sc},\varphi_{sc})$ depending on the depth of the layer near the front surface; s- polarization; For both figures $n=0.05, \Delta=0, \theta_{0}=17.5°, l_{x}=102, l_{y}=51, l_{z}=11.42$}
\end{center}
\par
\label{fig2}
\end{figure}

Note, however that attenuation of the curve shown in the Fig.2b is essentially slower than it can be expected if we suggest wave damping taking into account its propagation in two directions -- toward scattered atom and from them outside the medium. Fig.2b shows total contribution of atoms located inside layer with thickness $d$. From this contribution we can calculate partial contribution of atoms located in thin layer situated at the arbitrary depth $l$. Corresponding analysis shows that this partial contribution decreases approximately exponentially   $\exp(-\alpha l/\cos\theta_0)$ (the deviation from exponential dependence connects with mentioned above boundary effects). Index of exponential attenuation $\alpha$ is close to inverse mean free path of photon in the considered medium $\alpha \approx 1/l_{ph}$. So this index is two times smaller as compared with its expected value if we consider light propagation in two directions. This discrepancy connects with the fact that inside the atomic ensemble there is no coherent wave propagating in "backward" direction (of course, if the medium is semi-infinite or scattering from far edge can be neglected). Reflected coherent light beam exists only outside the atomic ensemble and this beam is a result of collective scattering by all atoms of the ensemble. Fig. 2 demonstrates that the determinative contribution to the reflection signal is given by subsurface layer with the depth comparable with wavelength.

\subsection{Comparison with Fresnel equations}

In this subsection we consider the dependence of the reflection coefficient  on the angle of incidence and show that under considered conditions such dependence cannot be described by Fresnel equations.

Comparing our results with Fresnel theory we have to take into account that the angular size of reflected light cone in our case is finite even for the plane incident wave because of finite sizes of the front surface of the atomic sample. For this reason it is naturally to determine the reflectivity $R$ as the ratio of the light power in the main maximum of angular distribution $P$ (see Fig.1) to the total power of light scattered in all directions $P_{0}$.

$P$ can be obtained as an integral of the angular distribution of scattered light $P(\theta_{sc},\varphi_{sc})$ over the reflected light cone $\Omega_{c}$.
\begin{eqnarray}%
P &=&\int\limits_{\Omega_{c}}P(\theta_{sc},\varphi_{sc})d\Omega
\label{4}
\end{eqnarray}%
$P_{0}$ can be obtained using the optical theorem.

Figure 3 shows the dependence of reflectivity on the angle of incidence. Two couples of curves are shown. The first one is calculated on the basis of microscopic approach and the second couple of curves are obtained from Fresnel equations.  Dielectric permittivity required for corresponding calculation were calculated by the method described in \cite{23},\cite{24}.
\begin{figure}
\begin{center}
{$\scalebox{0.4}{\includegraphics*{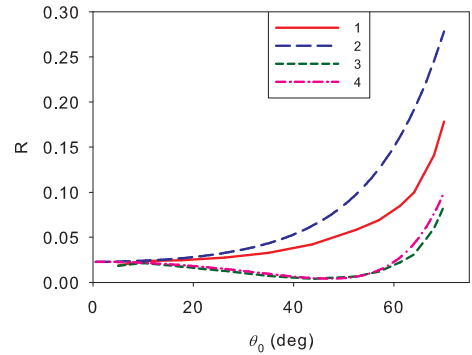}}$ }
\caption{(color online) Dependence of reflectivity on the angle of incidence; 1,2 s- polarization, 3,4 p-polarization; microscopic approach 1,3; Fresnel equations 2,4; the parameters of ensemble are the same as in Fig. 1}
\end{center}
\par
\label{fig3}
\end{figure}

Figure 3 demonstrates that for the atomic density $n=0.05$ we have a good agreement between quantum microscopic approach and Fresnel equations for p- polarization. However for s- polarizations there is noticeable quantitative disagreement between these two approaches. The situation changes more dramatically for bigger densities.

Figure 4 shows the angular dependence of reflectivity for the atomic density $n=0.5$. Resonant dipole-dipole interaction is so strong for this density that it causes negative real part of the dielectric permittivity in some spectral area \cite{23}-\cite{27}. For example, for the probe light with detuning $\Delta=\gamma_{0}$ the dielectric permittivity is equal to $\varepsilon=-0.125+1.542i$. The mean free path of photon corresponding to these parameters $l_{ph}=0.55$. Curves in the Fig. 4.a are calculated for this case. For comparison we add the plot corresponding to the negative detuning  $\Delta=-\gamma_{0}$ ($\varepsilon=1.80+1.40i$). The mean free path of photon for this detuning $l_{ph}=1.03$. Note that the dielectric permittivity was calculated in \cite{24} for spatially homogeneous (on average) atomic ensembles by the analysis of light propagation sufficiently far from their boundaries.

\begin{figure}
\begin{center}
{$\scalebox{0.4}{\includegraphics*{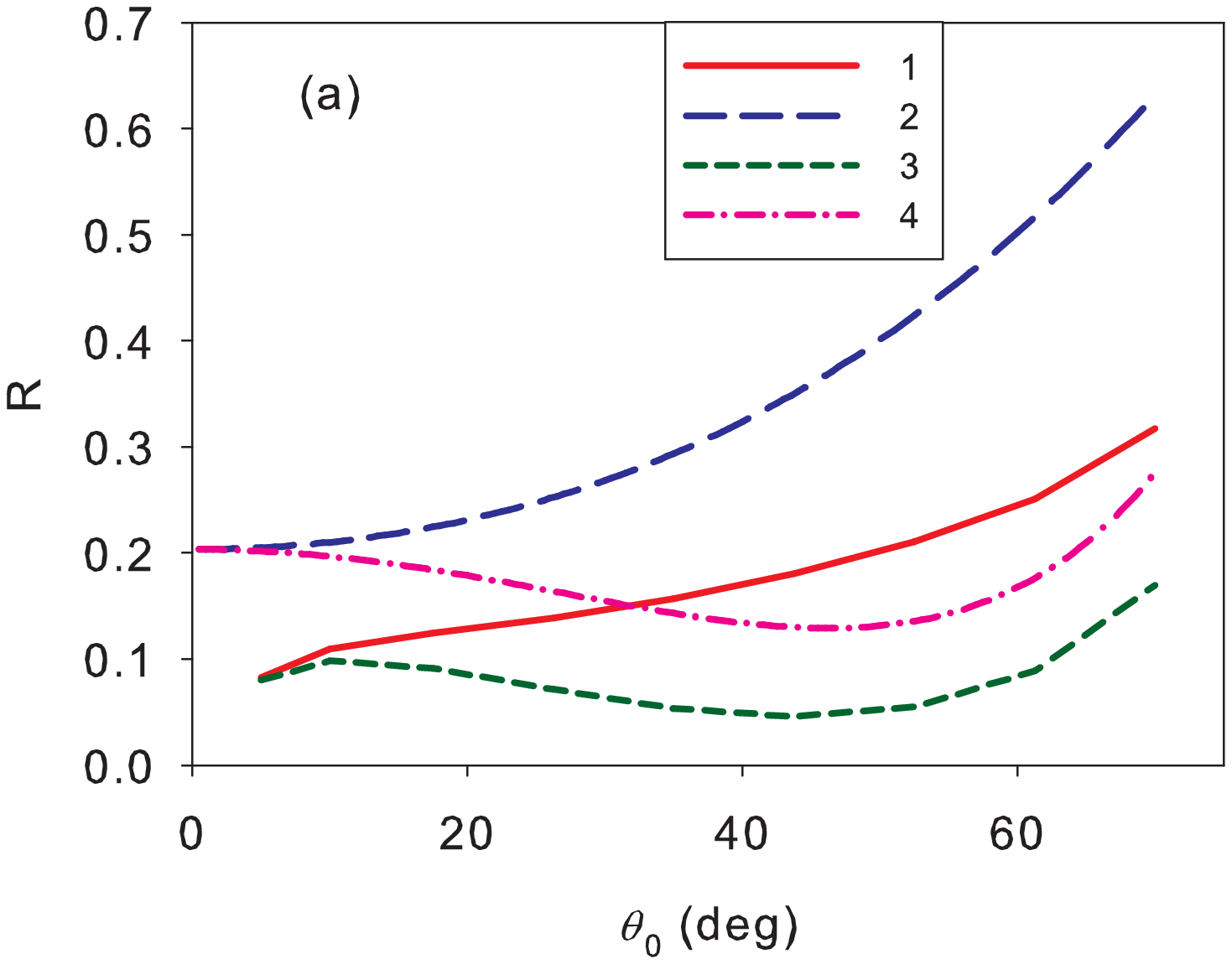}}$ }{$\scalebox{0.4}{%
\includegraphics*{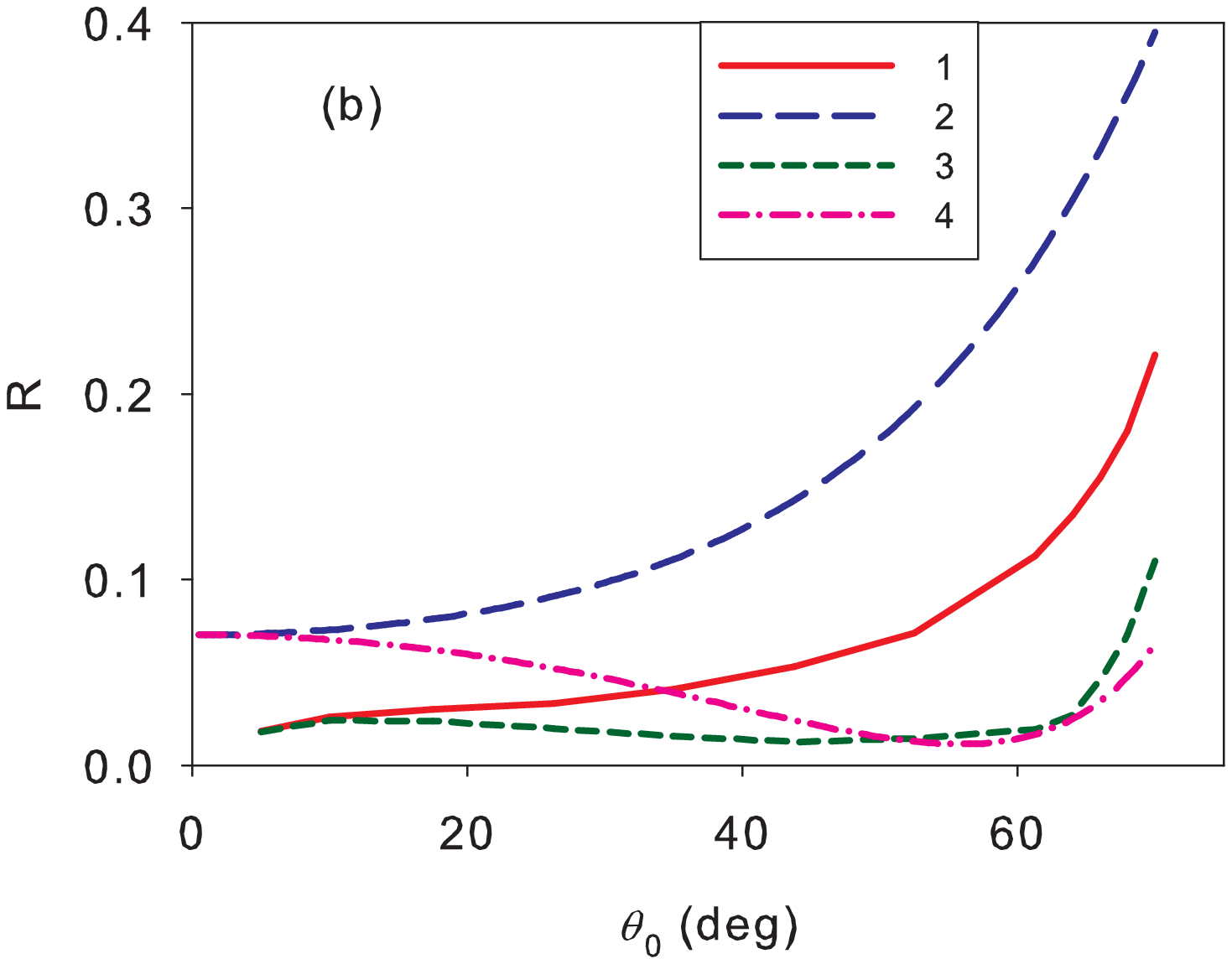}}$ }
\caption{(color online) Reflection coefficient depending on the angle of incidence, $n=0.5$; $\Delta=\gamma_{0}$ (a); $\Delta=-\gamma_{0}$ (b); 1,2 s- polarization, 3,4 p-polarization; microscopic approach 1,3; Fresnel equations 2,4}
\end{center}
\par
\label{fig4}
\end{figure}

From the Fig. 4 it is clearly seen that results obtained in the frame of microscopic approach differs essentially from predictions of Fresnel equations for both polarization channels  as well as for both considered detunings $\Delta=\gamma_{0}$ and $\Delta=-\gamma_{0}$.

In our opinion there are two main reasons for such discrepancy. First of all Fresnel equations require that averaged interatomic separations should be much less than light wavelength and photon mean free path in considered medium.  In our case it is not so. Both wavelength and mean free path of photon are comparable with inrteratomic separations. The second important peculiarity of considered physical conditions is essential role of boundary effects.  As we showed in the previous subsection under resonant reflection the main contribution into reflected signal is given by the surface layer which depth is about several photon mean free path. But just in this spatial domain the inhomogeneity in optical properties of a medium caused by the features of resonant dipole-dipole interaction is very essential \cite{25}. Atoms located in the surface layer responsible for reflected signal are in different physical condition as compared with atoms located inside the medium.

At the end of this section note that the important feature of quantum microscopic approach used in this work is that the resolvent matrix (\ref{1}) is determined numerically so it does not allow us to consider atomic ensemble with very big number of atoms. The calculations described in this paper were made for 2000 - 7000 atoms. In this regard, to make sure that observed results are not caused by small size of the cloud  we repeated for comparison our calculations for different sizes of the front surface of atomic ensemble. We increased the area of the front surface from 1.2 to 1.6 times and the difference in the reflection coefficient was at the level of computational error caused by statistical error of Monte-Carlo averaging mainly.

\section{Conclusion}
In this paper we analyze the reflection of quasi-resonant light from a plane surface of dense and disordered ensemble of motionless point scatters like impurity centers in solid. The calculation is performed on the basis of quantum microscopic approach. Solving the nonstationary Schrodinger equation for the joint system consisting of atoms and a weak electromagnetic field we calculate angular and polarization characteristics of light scattered by an ensemble in the form of a rectangular parallelepiped with big optical depth. The ratio between coherent and incoherent (diffuse) components of scattered light is also studied.

Microscopic approach allows us to analyze the influence of scatters located at different distances from the surface. This analysis shows that main contribution into reflected light comes from the surface layer which depth is determined by several mean free path of photon in considered medium. It proves that the inhomogeneity of dipole-dipole interaction near the surface essentially influences on the coherent reflection.

We studied the dependence of total reflected light power on the incidence angle for both s- and p- polarizations. The calculations are performed for different densities of scatters and different frequencies of a probe radiation. The reflection coefficient obtained in the framework of quantum microscopic approach is compared with Fresnel equations. It is shown that a disagreement between two approaches increases with atomic density. This discrepancy is explained by subsurface violation in the spatial homogeneity of the medium and by the fact that for resonant light the mean free path of photon is comparable with the average interatomic distance. It is shown that an important parameter here is $k_{0}l_{ph}$. A disagreement between two approaches increases with decrease of the value $k_{0}l_{ph}$.

We expect that observed disagreement between quantum microscopic approach and Fresnel equations (in case of resonant light) will be especially important for the case of light reflection from thin films. If the thickness of the film is comparable with the resonant wavelength we can consider the whole volume of the medium as the subsurface area. The approach employed in present work can be successfully used for this case even for films with inhomogeneous spatial distribution of atomic density. Furthermore, our approach allows to describe the light scattering by nanoclusters with a small number of atoms.

In our opinion, microscopic analysis of resonant reflection performed here will be useful for further improvement of optical detection methods based on coherent scattering of resonant light.

\section{Acknowledgements}
We acknowledge financial support from the Russian Foundation
for Basic Research (Grant No. RFBR-15-02-01013) and from the Ministry of Education
and Science of the Russian Federation (State Assignment
3.1446.2014K). A.S.K. appreciates also financial support from the RFBR (Grant No.14-02-31422), the Council for Grants of the President of the Russian Federation and the non-profit foundation "Dynasty".


\baselineskip20 pt

\begin{thebibliography}{99}
\bibitem{1} E. V. Timoshchenko, V. A. Yurevich, and Yu. V. Yurevich, Technical Physics
\textbf{58}, 251 (2013).

\bibitem{2} A. E. Kaplan and S. N. Volkov, Phys. Rev. Lett. \textbf{101}, 133902 (2008).

\bibitem{3} A. E. Kaplan and S. N. Volkov, Phys. Rev. A \textbf{81}, 043801 (2010).
\bibitem{4} A. V. Sel’kin, Yu. N. Lazareva, and V. A. Kosobukin, Journal of Optical Technology \textbf{78}, 519 (2011).
\bibitem{5} V. A. Sautenkov, H. Li, M. A. Gubin, Yu. V. Rostovtsev, and M. O. Scully, Laser Physics \textbf{21}, 153 (2011).

\bibitem{6} V. L. Velichanskii, R. G. Gamidov, G. T. Pack, and V. A. Sautenkov, JETP Lett. \textbf{52}, 136 (1990).

\bibitem{7} Ya. A. Fofanov, Quantum Electronics \textbf{39}, 585 (2009).

\bibitem{8} I. V. Zlodeev, Yu. F. Nasedkina, and D. I. Sementsov, Optics and Spectroscopy \textbf{113}, 208 (2012).

\bibitem{9} Yu. F. Nasedkina and D. I. Sementsov, Optics and Spectroscopy \textbf{104}, 591 (2008).

\bibitem{9a} G. Nienhuis, F. Schuller, and M. Ducloy, Phys. Rev. A \textbf{38}, 5197 (1988).

\bibitem{9b} V. A. Sautenkov, H. Lia, M. A. Gubin, Yu. V. Rostovtsev, and M. O. Scully, Laser Physics \textbf{21}, 153 (2011).

\bibitem{9c} Ya. A. Fofanov and A. A. Rodichkina, Optics and Spectroscopy \textbf{103}, 322 (2007).

\bibitem{9d} G. Nienhuis and F. Schuller, Phys. Rev. A \textbf{50} 50, 1586 (1994).

\bibitem{9e} J. Guo, J. Cooper, A. Gallagher, and M. Lewenstein, Optics Communications \textbf{110}, 732 (1994).

\bibitem{9f} H. Li, T. S. Varzhapetyan, V. A. Sautenkov, Y. V. Rostovtsev, H. Chen, D. Sarkisyan, M. O. Scully, Applied Physics B: Lasers and Optics \textbf{91}, 229 (2008).

\bibitem{9g} H. Li, V. A. Sautenkov, Y. V. Rostovtsev, and M. O. Scully, Journal of Physics B: Atomic, Molecular and Optical Physics \textbf{42}, 065203 (2009).

\bibitem{10} C. Cohen-Tannoudji, Nobel Lecture (1997).

\bibitem{11} E. Bimbard, R. Boddeda, N. Vitrant, A. Grankin, V. Parigi, J. Stanojevic, A. Ourjoumtsev, and P. Grangier, Phys. Rev. Lett. \textbf{112}, 033601 (2014).

\bibitem{12} D. Loss, D. P. Divincenzo, Phys. Rev. A \textbf{57}, 120 (1998).

\bibitem{13} K. A. Barantsev and A. N. Litvinov, J. Exp.
Theor. Phys. \textbf{118}, 569 (2014).

\bibitem{14} G. Wilpers,  T. Binnewies, C. Degenhardt, U. Sterr, J. Helmcke, and F. Riehle, Phys. Rev. Lett. \textbf{89}, 230801 (2002).

\bibitem{15} I. Courtillot, A. Quessada, and R. P. Kovacich, Phys. Rev. A \textbf{68}, 030501 (2003).

\bibitem{16} F. X. Esnault, N. Rossetto, D. Holleville, J. Delporte, and N. Dimarcq, Advances in Space Research \textbf{47}, 854 (2011).

\bibitem{17} D. Bouwmeester, A. Ekert, and A. Zeilinger, \emph{The Physics of
Quantum Information} (Springer-Verlag, Berlin, 2001).

\bibitem{18} J. S. Hodges, L. Li, M. Lu, E. H. Chen, M. E. Trusheim, S. Allegri, X. Yao, O. Gaathon, H. Bakhru, and D. Englund, New Journal of Physics \textbf{14}, 093004 (2012).

\bibitem{19} I. M. Sokolov, D. V. Kupriyanov, and M. D. Havey, J. Exp.
Theor. Phys.  \textbf{112}, 246 (2011).

\bibitem{20} I. M. Sokolov, A. S. Kuraptsev, D. V. Kupriyanov, M. D. Havey, and S. Balik, J. Mod. Opt. \textbf{60}, 50 (2013).

\bibitem{21} T. Ido, T. H. Loftus, M. M. Boyd, A. D. Ludlow, K. W. Holman, and J. Ye, Phys. Rev. Lett. \textbf{94}, 153001 (2005).

\bibitem{22} A. S. Kuraptsev and I. M. Sokolov, Phys. Rev. A \textbf{90}, 012511 (2014).

\bibitem{23} Ya. A. Fofanov, A. S. Kuraptsev, I. M. Sokolov, Opt. Spectrosc.
\textbf{112}, 401 (2012).

\bibitem{24} Ya. A. Fofanov, A. S.Kuraptsev, I. M. Sokolov, and M. D. Havey,
Phys. Rev. A \textbf{84}, 053811 (2011).
\bibitem{27} I. M. Sokolov, M. D. Kupriyanova, D. V. Kupriyanov, and M. D. Havey, Phys. Rev. A  \textbf{79}, 053405 (2009).

\bibitem{25} Ya. A. Fofanov, A. S. Kuraptsev, I. M. Sokolov, and M. D.
Havey, Phys. Rev. A \textbf{87}, 063839 (2013).
\bibitem{34} I. M. Sokolov, D. V. Kupriyanov, R. G. Olave and M. D. Havey, J. Mod. Opt. \textbf{57}, 1833 (2010).
\bibitem{35} S. E. Skipetrov and I. M. Sokolov, Phys. Rev. Lett. \textbf{112}, 023905 (2014).

\bibitem{26} L. Mandel and E. Wolf, \emph{Optical Coherence and Quantum Optics} (Cambridge University Press, Cambridge, 1995).

\bibitem{28} D. A. Varshalovich, A. N. Moskalev, and V. K. Khersonskiy, \emph{Quantum Theory of Angular Momentum} (World Scientific, Singapor, 1988)





\end{thebibliography}
\end{document}